\documentclass[twocolumn]{aastex62}

\newcommand{\ctext}[1]{\raise0.2ex\hbox{\textcircled{\scriptsize{#1}}}}

\received{June, 2022}
\revised{June, 2022}
\accepted{ApJ}

\shorttitle{WERGS. VII :  Redshift evolution of radio galaxy environments}
\shortauthors{Uchiyama et al.}

\begin{document}
\title{A Wide and Deep Exploration of Radio Galaxies with Subaru HSC (WERGS).  VII. \\
 Redshift Evolution of Radio Galaxy Environments at $z=0.3-1.4$}
\correspondingauthor{Hisakazu Uchiyama}
\email{uchiyama@cosmos.phys.sci.ehime-u.ac.jp}

\author{Hisakazu Uchiyama}
\affiliation{Research Center for Space and Cosmic Evolution, Ehime University, 2-5 Bunkyo-cho, Matsuyama, Ehime 790-8577, Japan} 

\author{Takuji Yamashita}
\affiliation{National Astronomical Observatory of Japan, Mitaka, Tokyo 181-8588, Japan} 
\affiliation{Research Center for Space and Cosmic Evolution, Ehime University, 2-5 Bunkyo-cho, Matsuyama, Ehime 790-8577, Japan} 

\author{Tohru Nagao}
\affiliation{Research Center for Space and Cosmic Evolution, Ehime University, 2-5 Bunkyo-cho, Matsuyama, Ehime 790-8577, Japan}

\author{Kohei Ichikawa}
\affil{Frontier Research Institute for Interdisciplinary Sciences, Tohoku University, Sendai 980-8578, Japan}
\affil{Astronomical Institute, Tohoku University, Aramaki, Aoba-ku, Sendai, Miyagi 980-8578, Japan}
\affil{Max-Planck-Institut f{\"u}r extraterrestrische Physik (MPE), Giessenbachstrasse 1, D-85748 Garching bei M{\"u}unchen, Germany}

\author{Yoshiki Toba}
\affiliation{National Astronomical Observatory of Japan, Mitaka, Tokyo 181-8588, Japan} 
\affiliation{Department of Astronomy, Kyoto University, Kitashirakawa-Oiwake-cho, Sakyo-ku, Kyoto 606-8502, Japan}
\affiliation{Academia Sinica Institute of Astronomy and Astrophysics, 11F of Astronomy-Mathematics Building, AS/NTU, No.1, Section 4, Roosevelt Road, Taipei 10617, Taiwan}
\affiliation{Research Center for Space and Cosmic Evolution, Ehime University, 2-5 Bunkyo-cho, Matsuyama, Ehime 790-8577, Japan} 

\author{Shogo Ishikawa} 
\affiliation{Center for Gravitational Physics, Yukawa Institute for Theoretical Physics, Kyoto University, Sakyo-ku, Kyoto 606-8502, Japan} 
\affiliation{National Astronomical Observatory of Japan, Mitaka, Tokyo 181-8588, Japan}

\author{Mariko Kubo}
\affiliation{Research Center for Space and Cosmic Evolution, Ehime University, 2-5 Bunkyo-cho, Matsuyama, Ehime 790-8577, Japan} 

\author{Masaru Kajisawa}
\affiliation{Research Center for Space and Cosmic Evolution, Ehime University, 2-5 Bunkyo-cho, Matsuyama, Ehime 790-8577, Japan} 

\author{Toshihiro Kawaguchi}
\affiliation{Department of Economics, Management and Information Science, Onomichi City University, Hisayamada 1600-2, Onomichi, Hiroshima 722-8506, Japan} 

\author{Nozomu Kawakatu}
\affiliation{National Institute of Technology, Kure College, 2-2-11, Agaminami, Kure, Hiroshima 737-8506, Japan}

\author{Chien-Hsiu Lee} 
\affiliation{NSF's National Optical-Infrared Astronomy Research Laboratory, Tucson, AZ 85742, USA} 

\author{Akatoki Noboriguchi}
\affiliation{School of General Education, Shinshu University, 3-1-1 Asahi, Matsumoto, Nagano 390-8621, Japan}

\begin{abstract} 
We examine the redshift evolution of density environments around 2,163 radio galaxies with the stellar masses of $\sim10^{9}-10^{12} M_\odot$ between redshifts of $z=0.3-1.4$, based on the Hyper Suprime-Cam Subaru Strategic Program (HSC-SSP) and Faint Images of the Radio Sky at Twenty-cm (FIRST). 
We use the $k$-nearest neighbor method to measure the local galaxy number density around our radio galaxy sample. 
We find that the overdensities of the radio galaxies are weakly but significantly anti-correlated with redshift. 
This is consistent with the known result that the relative abundance of less-massive radio galaxies increases with redshift, because less-massive radio galaxies reside in relatively low density regions. 
Massive radio galaxies with stellar mass of $M_* >10^{11}M_\odot$ are found in high density environments compared with the control sample galaxies with radio-non-detection and matched-stellar-mass. 
 Less-massive radio galaxies with $M_* <10^{11}M_\odot$ reside in average density environments. 
The fraction of the radio galaxies associated with the neighbors within a typical major merger scale, $<70$ kpc,  is higher than (comparable to) that of the control galaxies at $M_* >10^{11}M_\odot$ ($M_* <10^{11}M_\odot$). 
We also find that the local densities around the radio galaxies are anti-correlated with the radio luminosities and black hole mass accretion rates at fixed stellar mass. 
These findings suggest that massive radio galaxies have matured through galaxy mergers in the past, and have supermassive black holes  whose mass accretion almost ceased at $z>1.4$, while less-massive radio galaxies undergo active accretion just at this epoch, as they have avoided such merger events. 
\end{abstract}


\section{INTRODUCTION}  

Radio galaxies host active galactic nuclei (AGNs) that launch strong radio jets/lobes, which are expected to affect the star formation not only in the host halo but also the surrounding halos \citep[][]{Morganti05, McNamara07, Birzan08, Shabala11, Yuan16,  Nesvadba17}. 
Thus,  in order to understand galaxy formation and evolution throughout the history of the universe, it is key to understand how/where radio galaxies appear. 

In the local Universe, radio galaxies are often found in rich environments, such as galaxy overdense regions or massive halos \citep[e.g.,][]{Peacock91, Mag04, Venturi07}. 
This fact is qualitatively consistent with radio jet/lobe triggering mechanism, Blandford Znajek process \citep[][]{Blandford77}. 
According to this process, the radio jet/lobe luminosity is a monotonically increasing function of black hole mass and spin which are built up by galaxy mergers  \citep[e.g., ][]{Fanidakis11}. 
Thus, radio galaxies are likely to easily appear in galaxy overdense regions, where galaxy mergers had been frequently experienced.

The galaxy overdensities around radio galaxies depend on their types significantly \citep[][]{Ramos13,Ching17}.  
Radio galaxies can be divided into two types, high-excitation radio galaxies (HERGs) and low-excitation radio galaxies (LERGs). 
HERGs have strong radiation from accretion disk, caused by quasar mode accretion \citep[][]{Bower06, Croton06}. 
This accretion channel dominantly occurs in the dark matter halos whose masses are $\sim10^{12} h^{-1} M_\odot$ \citep[][]{Orsi16}. 
On the other hand, LERGs appear through relatively slow mass accretion onto supermassive black hole (SMBH), and are hosted by more massive halos than HERGs \citep[][]{Turner15, Orsi16}. 
At $z<0.4$, LERGs are observed to reside in galaxy denser regions than HERGs significantly \citep[][]{Ramos13,Ching17}. 
HERGs are also reported to have lower stellar masses than LERGs on average \citep[][]{Ching17}. 

The overdensities around radio galaxies are expected to depend on the age of the universe.  
The relative abundance of HERGs in radio galaxies is observed to increase with redshift in the range of $z=0.4-0.8$ \citep[][]{Donoso09}. 
The abundance of radio AGNs hosted by star forming galaxies or less massive ($M_*<10^{11} M_\odot$) galaxies is reported to rise with redshift \citep[][]{Delvecchio18}.  
\citet{Delvecchio18} found that the radiatively-efficient accretion in radio AGN becomes to be dominant at $z\gtrsim1$, suggesting the increase of the proportion of less-massive HERGs at this epoch. 
These facts imply that radio galaxies do not reside in the most overdense regions, on average, at $z\gtrsim1$. 
Thus, it is key to examine the redshift evolution of the galaxy densities around radio galaxies over the wide stellar mass range, up to $z\gtrsim1$. 
Unfortunately, the redshift evolution of the radio galaxy environments are still shrouded in a deep fog. 
\citet{Kolwa19} found that radio AGN, on average, reside in the overdense regions at $z<0.8$ by using the data of $1-2$ GHz Very Large Array (VLA) survey and the Sloan Digital Sky Survey (SDSS) Stripe 82 \citep[][]{Heywood16}. 
That wide coverage over $\sim100$ deg$^2$ allows us to characterize the radio galaxy environments statistically. 
However, the radio AGN sample and also the surrounding galaxy sample is biased toward massive galaxies ($M_*>10^{11}M_\odot$) due to the shallowness of the survey ($g<24.5$ and $K\lesssim18.2$).  
\citet{Malavasi15} found that radio AGN always are located in rich and dense environments up to $z\sim2$ in the Cosmological Evolution Survey (COSMOS) field whose survey area is $\sim2$ deg$^2$. 
This small area survey could miss rare objects and make a sparse redshift distribution that is insufficient to investigate the redshift evolution.  

The data of a Wide and Deep Exploration of Radio Galaxies with Subaru HSC \citep[WERGS, ][]{Yamashita18,Toba19, Yamashita20, Ichikawa21, Uchiyama21} can  overcome the difficulties. 
WERGS is the wide optical counterpart survey of radio galaxies with the optical depth down to $i\sim26$.  
This project is based on the  wide and deep optical imaging data \citep[][]{Aihara18a} produced by Subaru/Hyper Suprime-Cam Subaru Strategic Program (HSC-SSP) survey and the 1.4 GHz radio continuum catalog of the Faint Images of the Radio Sky at Twenty-cm survey using VLA \citep[``FIRST''; ][]{Becker95, Helfand15}. 
The combination of these survey can make the statistical radio galaxy sample with  a wide stellar mass range of $\sim10^9 - 10^{12}  M_\odot$ at $z<1.4$.   

In this paper, we examine the redshift evolution of radio galaxy environments up to $z=1.4$, by using the data of WERGS. 
The radio galaxy sample is extracted from WERGS data. 
We also construct the control galaxy sample covering similar stellar masses and redshifts, but without radio detections. 
We can refer to the possible relation between the density environments and the radio jets, by comparing the radio galaxies with the control ones. 
The dependency of the density environments on their stellar masses are also investigated. 

The paper is organized as follows. 
In section 2, we describe the data of HSC-SSP, FIRST and WERGS. 
The construction of the radio galaxy and control galaxy samples is also described. 
In section 3, the method to measure the local densities around radio and control galaxies is explained. 
The results of the redshift evolution of the radio galaxy environments and the possible dependency of the local densities on the stellar masses of the galaxies are shown in section 4. 
The implications of our results are discussed in section 5. 
Finally, in section 6 we summarize our findings. 
We assume the following cosmological parameters: 
$\Omega_{M} = 0.27 $, $\Omega_{\Lambda} = 0.73$, $H_{0} = 70 ~ $km s$^{-1}$ Mpc$^{-1}$,  
and magnitudes are given in the AB system. 

\section{DATA AND SAMPLE} 

\subsection{Data}

The Subaru HSC-SSP survey \citep[][]{Aihara18b,Kawanomoto17,Komiyama18,Furusawa18,Bosch18,Ivezic08, Axelrod10, Juric15,Schlafly12, Tonry12, Magnier13} is an unprecedented deep-and-wide optical survey using HSC with 116 2K $\times$ 4K Hamamatsu fully-depleted CCDs and the field-of-view of 1.$^\circ$5 diameter \citep[][]{Miyazaki12, Miyazaki18}. 
In this present study,  we use the ``wide layer" data of DR S16A \citep[][]{Aihara18a}, which consists of wide field image of $>200$ deg$^2$ with a median seeing of $0.\arcsec6 - 0.\arcsec8$ and is taken by the five optical filters of $grizy$-bands.  
The $5\sigma$ limiting magnitudes of $grizy$-bands  for point sources measured in $2.0~ $ arcsec apertures are expected to be $26.5$, $26.1$, $25.9$, $25.1$, and $24.4$, respectively. 
We use cModel magnitudes, which are measured by fitting two components that are PSF-convolved galaxy models (de Vaucouleurs and exponential) to the source profile \citep[]{Abazajian04}.


The survey area of VLA FIRST is $\sim10,000$ deg$^2$, and cover the HSC-SSP survey area. 
The images produced  have a typical root-mean-square photometric noise of 0.15 mJy. 
The angular resolution of the FIRST survey is 5.4 arcsec.

\subsection{Sample}
\subsubsection{WERGS}
We describe the key steps of the construction of the WERGS radio source catalog  \citep[][]{Yamashita18}. 

The HSC-SSP sources were extracted from the wide layer data of HSC-SSP. 
The edge of each independent field was masked because the limiting magnitudes at the edges are relatively shallower than those in the central regions. 
The HSC-SSP flags summarized in \citet{Yamashita18} were imposed for the sources to remove fake detections. 
The signal to noise ratio ($SNR$) for the $riz$-bands were required to be greater than five. 
As a result, 23,795,523 sources were found in the effective area of 154 deg$^2$ (hereinafter referred to as ``HSC sources"). 

The HSC-FIRST sample was constructed by matching the HSC sources with the FIRST sources. 
The FIRST sources with $>1$ mJy radio flux and the $P(S)$ of $<0.05$ were extracted from the final release FIRST catalog \citep[][]{Helfand15}. 
Here, $P(S)$ is a probability that the source is a side lobe of a nearby bright source.  
7,072 FIRST radio sources were found in the HSC-SSP wide layer. 
Then, the radio sources were matched with HSC sources using a search radius of $1.0$ arcsec, which is the separation where the contamination rate and  completeness function intersect. 
The contamination rate and the completeness were estimated to be $14$ \% and $93$ \%, respectively. 
As a result,  3,579 HSC-FIRST sources were found in the wide layer \citep[][]{Yamashita18}. 

\subsubsection{Radio galaxies} 

We extract radio galaxies from the HSC-FIRST sources using the rest-frame 1.4 GHz luminosities, $L_{1.4\text{GHz}}$ [W Hz$^{-1}$] according to \citet[][]{Ichikawa21}. 
The $L_{1.4\text{GHz}}$ of the HSC-FIRST sources were calculated to be $\sim10^{22.0-27.5}$ W Hz$^{-1}$  assuming a radio spectral index of  $\alpha = -0.7$ with  the form of $f_\nu \propto \nu^\alpha$ \citep[][]{Yamashita18}. 
The redshift of each source is calculated using the photo-$z$ calculated by Mizuki spectral energy distribution (SED) fitting code \citep[][]{Tanaka18}. 
We use only the HSC-FIRST sources with $L_{1.4\text{GHz}}>10^{24}$ W Hz$^{-1}$. 
This luminosity cut can effectively split radio galaxies and star forming galaxies \citep[e.g.,][]{Condon13}. 
 Point sources are excluded as quasars using the same method as \citet{Yamashita18}. 
This reduces the number of the sources to 2,918. 

\begin{figure}
\begin{center}
\includegraphics[width=1.0\linewidth]{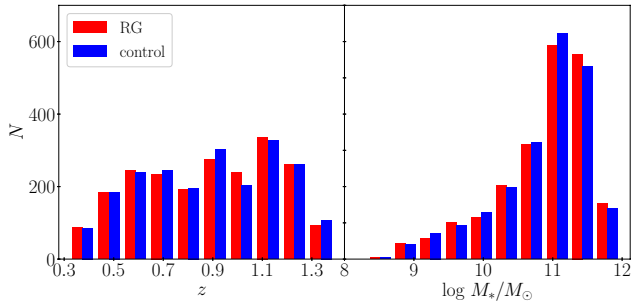}
\end{center}
\caption{Distributions of the redshifts and stellar masses of the radio galaxy (red bars) and control galaxy (blue bars) samples.  The blue bars are shifted by +0.05 relative to the red bars for visibility. 
}\label{rgcontrol}
\end{figure}

At least one of the HSC filters can detect the flux at the bluer or redder side of the Balmer break wavelength of the sources at $z=0.3-1.4$ \citep[e.g.,][]{Ishikawa20}. 
For the photo-$z$ accuracy,  
we select galaxies with photo-$z$ error  $\sigma_{z} <0.1 (1+z)$ and reduced $\chi$ square $\chi^2_\nu <3$,  based on \citet{Yamashita18}, \citet{Toba19}, and \citet{Ichikawa21}. 
As a result, we obtain  2,163 radio galaxy candidates at $z=0.3-1.4$ (hereinafter referred to as  ``radio galaxies"). 
The radio galaxies have stellar masses, $M_*$ [$M_\odot$], of $\sim10^{9-12} M_\odot$ estimated in the Mizuki SED fitting code. 
The redshift and stellar mass distributions of the radio galaxies are shown in Figure \ref{rgcontrol}. 

\subsubsection{$z$ \& $M_*$ matched control sample} 
A control galaxy sample without radio emission is constructed to compare the environments at between radio galaxies and radio-quiet galaxies. 
First, we impose the same photo-$z$ conditions as the radio galaxies on the HSC sources.  
Then, we extract HSC sources (hereinafter, ``control galaxies"") that have the same redshift and stellar mass as each radio galaxy but have no detection in FIRST. 
The absolute values of the difference in redshift and stellar mass between a radio galaxy and a corresponding control galaxy, $\Delta z$ and $\Delta M_*$ [$M_\odot$], obey the conditions $\Delta z < \sigma^{\mathrm{rg}}_{z}$ and $\Delta M_* < \sigma^{\mathrm{rg}}_{M_*}$, respectively. 
$\sigma^{\mathrm{rg}}_{z}$ and $\sigma^{\mathrm{rg}}_{M_*}$ [$M_\odot$] are the errors of the redshift and stellar mass of the radio galaxy, respectively. 
If multiple control galaxies are selected for a radio galaxy, one control galaxy is randomly selected. 
We confirm that there is at least one corresponding control galaxy for each radio galaxy. 
As a result, 2,163 control galaxies are selected. 

Figure \ref{rgcontrol} shows the stellar mass and redshift distributions of the control galaxies. 
There is no difference in the distributions between the radio and control galaxy samples statistically. 
The $D$- and $P$-values for the redshift (stellar mass) distribution between the radio and control galaxies are $0.03$ (0.03) and $0.51$ (0.18) in the Kolmogorov-Smirnov test, respectively.

\subsubsection{Galaxies for density measurements} 
We construct a galaxy sample to be used in measuring the density environments around the radio and control galaxies.  
The same photo-$z$ conditions are imposed on the HSC sources as on the radio galaxies. 
We use photo-$z$ sources with the stellar masses of $>10^{10.1}M_\odot$. 
The stellar mass cut corresponds to the $70$ \% completeness limit of the stellar mass functions at $z=1.4$  in the Cosmic Evolution Survey (COSMOS) field \citep[][]{Ishikawa20}. 
As a result, we obtain 2,080,217 galaxies at $z=0.3-1.4$ for density measurement (hereinafter referred to as ``density galaxies").

\begin{deluxetable}{@{\extracolsep{4pt}}llrlrlrlrlr}
\tablecaption{Spearman rank correlation test for overdensity and redshift. \label{t_redshift}}
\tablecolumns{9}
\tablewidth{0pt}
\tablehead{
\colhead{}                     & \multicolumn{2}{c}{radio galaxy}                       &\multicolumn{2}{c}{control galaxy}                 \\
\cline{2-3}  \cline{4-5} 
\colhead{overdensity}    & \colhead{$\rho$ \tablenotemark{a}}     & \colhead{$P$\tablenotemark{b}}  & \colhead{$\rho$ \tablenotemark{a}}     & \colhead{$P$\tablenotemark{b}} 
}
\startdata 
$1+\delta^{k=1}$          & $-0.071$             & $1.3\times10^{-3}$       &$5.6\times10^{-3}$                            &0.80       \\ 
$1+\delta^{k=2}$          &$-0.11$        &   $1.6\times10^{-7}$    &$-1.0\times10^{-3}$                            &0.96        \\ 
$1+\delta^{k=5}$          &$-0.10$                           & $3.1\times10^{-6}$                    &$5.5\times10^{-3}$                            &0.81         \\          
\enddata
\tablenotetext{a}{Correlation coefficient of Spearman rank correlation test. }
\tablenotetext{b}{$P$-value of Spearman rank correlation test. }
\end{deluxetable}

\begin{deluxetable*}{@{\extracolsep{4pt}}llrlrlrlrlrlrlrlrlr}
\tablecaption{Spearman rank correlation test for overdensity and stellar mass at each redshift bin. \label{t_stellarmass_bin}}
\tablecolumns{9}
\tablewidth{0pt}
\tablehead{
\colhead{}   & \multicolumn{2}{c}{$z=0.3-0.5$} & \multicolumn{2}{c}{$z=0.5-0.8$} & \multicolumn{2}{c}{$z=0.8-1.1$}& \multicolumn{2}{c}{$z=1.1-1.4$}   \\
\colhead{overdensity}    & \colhead{$\rho$ \tablenotemark{a}}     & \colhead{$P$\tablenotemark{b}}  & \colhead{$\rho$\tablenotemark{a}}     & \colhead{$P$\tablenotemark{b}} & \colhead{$\rho$ \tablenotemark{a}}     & \colhead{$P$\tablenotemark{b}}  & \colhead{$\rho$\tablenotemark{a}}     & \colhead{$P$\tablenotemark{b}} 
}
\startdata 
radio galaxy                  \\
\hline
$1+\delta^{k=1}$          &0.24                           & $1.0\times10^{-3}$                  &0.31                           &$3.7\times10^{-15}$   &0.30                           & $5.0\times10^{-15}$                  &0.27                           &$1.1\times10^{-11}$         \\ 
$1+\delta^{k=2}$          &0.21                           & $2.8\times10^{-3}$                  &0.34                           &$7.8\times10^{-19}$   &0.29                           & $5.3\times10^{-15}$                  &0.22                           &$3.0\times10^{-8}$          \\ 
$1+\delta^{k=5}$         &0.24                          & $9.5\times10^{-4}$                  &0.29                           &$4.2\times10^{-14}$   &0.28                           & $1.3\times10^{-13}$                  &0.21                           &$1.7\times10^{-7}$          \\          
\hline
control galaxy           \\
\hline 
$1+\delta^{k=1}$          &0.16                           & 0.032                  &0.074                           &0.061   &0.093                           & 0.016                  &0.14                           &$1.0\times10^{-3}$ \\
$1+\delta^{k=2}$          &0.13           & 0.078                  &0.11              &$6.0\times10^{-3}$   &0.11                           & $3.0\times10^{-3}$                  &0.16                &$1.0\times10^{-4}$     \\
$1+\delta^{k=5}$          &0.077        & 0.30                  &0.094                           &0.017   &0.12                           & $1.5\times10^{-3}$                  &0.16                           &$1.2\times10^{-3}$ 
\enddata
\tablenotetext{a}{Correlation coefficient of Spearman rank correlation test. }
\tablenotetext{b}{$P$-value of Spearman rank correlation test. }
\end{deluxetable*}

\subsection{Radio galaxy with IR photometry} 
IR photometry for our radio sample is taken from \citet{Toba19}.  
The CIGALE \citep[Code Investigating GALaxy Emission;][]{Burgarella05, Noll09, Boquien19} SED fitting with the stellar, star formation, AGN, and radio components was conducted for the HSC sources  in \citet{Toba19}. 
The IR data used in the SED fitting  comprises of data from the VISTA Kilo-degree Infrared Galaxy Survey \citep[VIKING; ][]{Arnaboldi07}, the Wide-field Infrared Survey Explorer \citep[WISE; ][]{Wright10}, and the Herschel Space Observatory \citep[][]{Pilbratt10} Astrophysical Terahertz Large Area Survey \citep[H-ATLAS; ][]{Eales10}. 
In the SED fitting, the photometric redshift estimated by photo-$z$ Mizuki code \citep[][]{Tanaka18} was used.


The physical quantities we use in this study are obtained by transforming the quantities obtained from the SED fitting. 
The bolometric AGN luminosity, $L_{\mathrm{AGN}}$ [$L_\odot$], can be obtained through the bolometric correction for the IR luminosity $L_{\mathrm{IR}}$ [$L_\odot$] obtained by the SED fitting; 
$L_{\mathrm{AGN}}=3.0\times L_{\mathrm{IR}}$ \citep[][]{Delvecchio14, Inayoshi18, Ichikawa21}.  
The specific black hole mass accretion rate ($sBHAR$ [erg s$^{-1}$ $M_\odot^{-1}$]) is defined as $L_{\mathrm{AGN}}/M_*$ \citep[e.g., ][]{Mullaney12, Ichikawa21}. 
$sBHAR$ is good indicator of the Eddington ratio  $\lambda_{\mathrm{Edd}}$ ($\propto L_{\mathrm{AGN}}/M_{\mathrm{BH}}$) according to the $M-\sigma$ relation \citep[][]{Magorrian98, Marconi03}. 
According to \citet{Ichikawa21}, $sBHAR$ can be expressed by a function of $M_*$ and $\lambda_{\mathrm{Edd}}$: 
\begin{equation}
sBHAR = 4.6 \times 10^{35} \lambda_{\mathrm{Edd}} \left(\frac{M_*}{10^{10} M_\odot}\right)^{0.4}, \label{sbhar_eq}
\end{equation}
assuming $M_{\mathrm{BH}} \propto M_*^{1.4}$ \citep[][]{Kormendy13}. 
The number of the radio galaxies with $L_{\mathrm{AGN}}$ and $sBHAR$ is found to be 1257 (hereinafter, we call the subsample to ``IR radio galaxies"). 

\section{Measurement of galaxy environment}
We use the $k$-nearest neighbor method to measure the local density environments of the radio galaxies and  control galaxies. 
The details of the method are as follows.

 \begin{figure*}
\begin{center}
\includegraphics[width=0.9\linewidth]{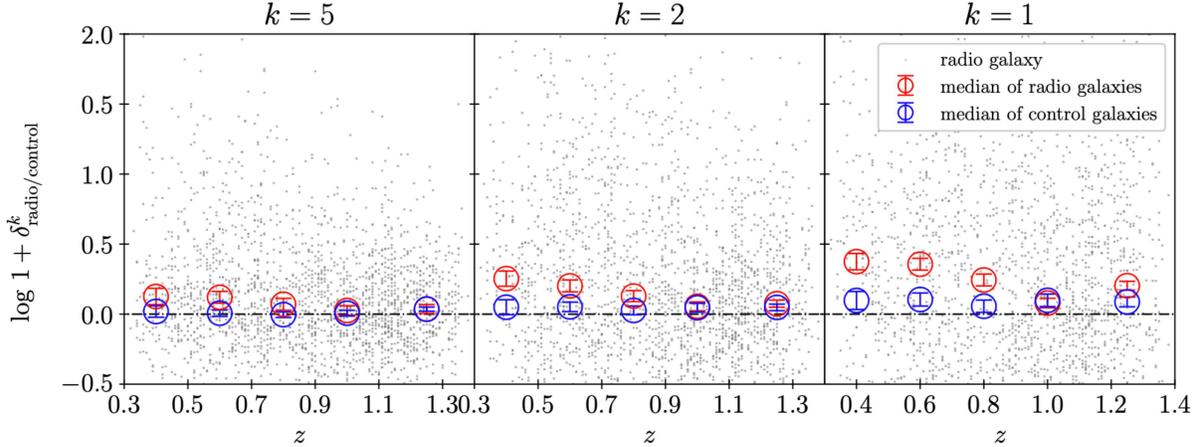}
\end{center}
\caption{The redshift evolution of the overdensities of the radio galaxies and control galaxies.  Left, middle, and right panels show the overdensities with $k=5$, 2 and 1, respectively. The median and the standard error of median of the overdensities of the radio (control)  galaxies in each redshift bin are shown by the red (blue) open circle and error bar, respectively. 
The grey dots indicate the overdensities of the radio galaxies. 
}\label{redshift}
\end{figure*}

The local density of a radio/control galaxy, $\Sigma_{\mathrm{radio/control}}^{k}$ [pkpc$^{-2}$], is estimated through the following equation.  
\begin{equation} 
\Sigma_{\mathrm{radio/control}}^{k} \equiv \frac{k+1}{\pi d_k^2} ~~ \mathrm{pkpc}^{-2}, \label{eq;01}
\end{equation} 
where, $d_k$ [pkpc] is the projected distance from a radio/control galaxy to the $k$-th nearest density galaxy within the redshift range of [$z-\sigma_z, z+\sigma_z$] \citep[][]{Lai16}, 
where $z$ and $\sigma_z$ are the redshift and its error of the radio/control galaxy, respectively. 
\citet{Lai16} demonstrated that when the photometric redshift error of a galaxy is employed as a redshift slice, the measured overdensity of a galaxy sample with photometric redshifts can best trace the true one. 
In the calculation of the local density (\ref{eq;01}), we correct it by the fraction of the non-masked projected area within a circle around a radio/control galaxy with a radius $d_k$ [pkpc]. 
Then, the overdensity ($1 + \delta^k_{\mathrm{radio/control}}$) for a radio/control galaxy is defined by 
\begin{equation}
1 + \delta^k_{\mathrm{radio/control}} \equiv \frac{\Sigma_{\mathrm{radio/control}}^{k}}{\left\langle\Sigma_{\mathrm{field}}^{k}\right\rangle},   \label{eq;2}
\end{equation}
where, $\left\langle \Sigma_{\mathrm{field}}^{k} \right\rangle$  [pkpc$^{-2}$] is the median local density of photo-$z$ galaxies within the same redshift range as the radio/control galaxy. 
If $1+\delta^{k}_{\mathrm{radio}}$ is higher than unity, the density around radio galaxies is overdense compared to field. 
We choose the overdensities with $k=1$, 2, and $5$ to compare the local densities in between different physical scales. 
By using the defined local densities, we examine their redshift evolution up to $z=1.4$. 


The medians of the projected distances $d_k$ are about $212$, $375$, and $697$ pkpc for $k=1$, 2 and $5$ for the radio galaxy sample, respectively. 
It is important to compare the projected distance to the nearest neighbor, $d_{k=1}$, with the typical major merger scale $<70$ pkpc \citep[][]{Larson16}. 
This comparison could allow us to understand whether the triggering of radio galaxies is associated with major mergers or not (section 5.1).

\begin{figure*}
\begin{center}
\includegraphics[width=1.0\linewidth]{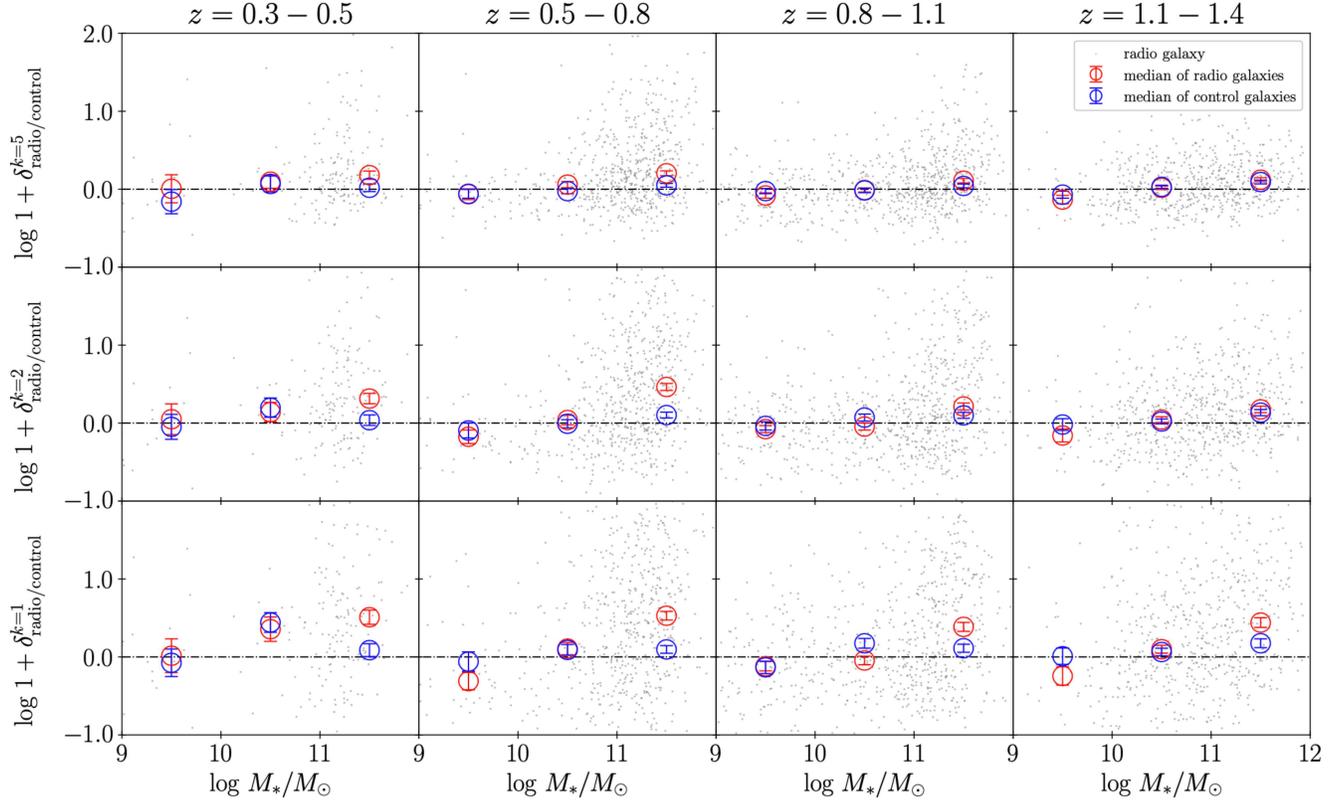}
\end{center}
\caption{The dependence of the overdensities of the radio/control galaxies on the  stellar masses at $z=0.3-0.5$, $z=0.5-0.8$, $z=0.8-1.1$, and $z=1.1-1.4$ from left to right.  
From the top row to the bottom,  the overdensities with $k=5$, 2 and 1 of the radio and control galaxies are shown. The median and the standard error of median of the overdensities of the radio (control) galaxies in each logarithmic stellar mass bin are shown by the red (blue)  open circle and error bar, respectively. 
The grey dots show the overdensities of the radio galaxies. 
}\label{stellarmass_bin}
\end{figure*}

\section{RESULTS} 
\subsection{Redshift evolution of density environments} 
We show the measured overdensities of the radio galaxy sample and the control galaxy one as a function of redshift in Figure \ref{redshift}.
We find that at $z<0.9$ the median $1+\delta^{k}_{\mathrm{radio}}$ for all $k$ is significantly higher than unity. 
The excess decreases as $k$ increases. 
These suggest that the radio galaxies at $z<0.9$, on average, reside in galaxy overdense compact regions. 
On the other hand, at $z>0.9$, the  median overdensities of the radio galaxies are close to unity for all $k$. 
These results suggest that the ambient densities of the radio galaxies are negatively correlated with redshift. 
The Spearman rank correlation test supports this negative correlation. 
The results of the test are summarized in Table \ref{t_redshift}. 

For the stellar-mass-matched control sample,  the overdensities for all $k$ are confirmed to be, on average, around unity over all the redshifts that we investigate. 
This suggests that the radio jet launches are related with the galaxy overdensities. 


\subsection{Dependence of the overdensities on the stellar masses}  

The dependence of the overdensities of the radio galaxies on their stellar masses is shown in Figure \ref{stellarmass_bin}. 
We divide the radio/control galaxies into four subsamples with $z = 0.3-0.5$, $0.5-0.8$, $0.8-1.1$, and $1.1-1.4$, in order to exclude the redshift dependency on the overdensities.  
At the low stellar mass regime (log $M_*/M_\odot < 11$), the median overdensities of the radio and control galaxies are comparable, and are around unity for all $k$. 
The control galaxies reside in the high density regions in the high stellar mass regime of log $M_*/M_\odot > 11$. 
The ambient densities of radio galaxies are, on average, significantly higher than those of the control galaxies  at log $M_*/M_\odot > 11$ for the case of $k=1$. 
The positive correlations between stellar mass and density are found to be significant using the Spearman rank correlation test as summarized in the Table \ref{t_stellarmass_bin}. 
The overdensities of the radio galaxies are more strongly correlated with the stellar masses compared to the control ones. 
This overdensity excess decreases for $k=2$ and $5$. 
There are no significant differences of the overdensities  $k=2$ and $5$ between the radio and control galaxies at $z>0.8$. 

We can compare our findings with previous studies. 
\citet{Kolwa19} found that massive radio AGNs, on average, reside in the kpc-scale overdense regions.  
\citet{Malavasi15} studied the environments of the radio AGNs with $M_*>10^{10} M_\odot$ by the deep but narrow survey in the COSMOS field and 
found that the radio AGNs reside  in high density regions compared to the non-radio sources with the same stellar masses. 
These results are consistent with our result of the massive radio galaxies. 



\section{Discussion}




\subsection{Redshift evolution of radio galaxy environments} 
We found the weak and negative correlation of the overdensity of the radio galaxies with redshift at $z = 0.3 - 1.4$.
At $z<0.9$, the radio galaxies on average reside in the galaxy overdense regions, while at $z>0.9$, the overdensities of the radio galaxies approach to be around unity. 
\citet{Kolwa19} found that at $z<0.8$, radio AGNs exist in the overdense regions. 
We, for the first time, have given a statistical insight that the overdensities of radio galaxies tend to decrease with redshift at $z>0.9$. 

This can be explained by the increase of the relative abundance of less-massive HERGs with redshift, assuming that the local observational result that the less-massive HERGs reside in the relatively low dense regions compared to massive LERGs \citep[][]{Ching17} is valid at $z=0.3-1.4$. 
In fact, \citet{Donoso10} found that the proportion of less-massive HERGs increases with redshift. 
\citet{Delvecchio18} also suggested that at $z\gtrsim1$ the accretion disk of radio AGN tend to be radiatively efficient. 
This means the increase of the population of HERGs in radio galaxies beyond $z\sim1$.  
In our WERGS radio galaxy sample, the radio galaxies with high $sBHAR$, that is, HERG-like galaxies are dominantly distributed at $z\gtrsim1$ \citep[][]{Toba19, Ichikawa21}.  
The stellar masses of the radio galaxies are also found to be significantly anti-correlated with redshift. 
The correlation coefficient, $\rho$, and $P$-value in the Spearman rank correlation test are $-0.17$ and $1.2\times10^{-9}$, respectively. 
These results suggest that the relative abundance of less-massive HERGs in the radio galaxy sample is high at $z\gtrsim1$, and thus, the overdensities of the radio galaxies tend to decrease with redshift.

\begin{figure*}
\begin{center}
\includegraphics[width=0.75\linewidth]{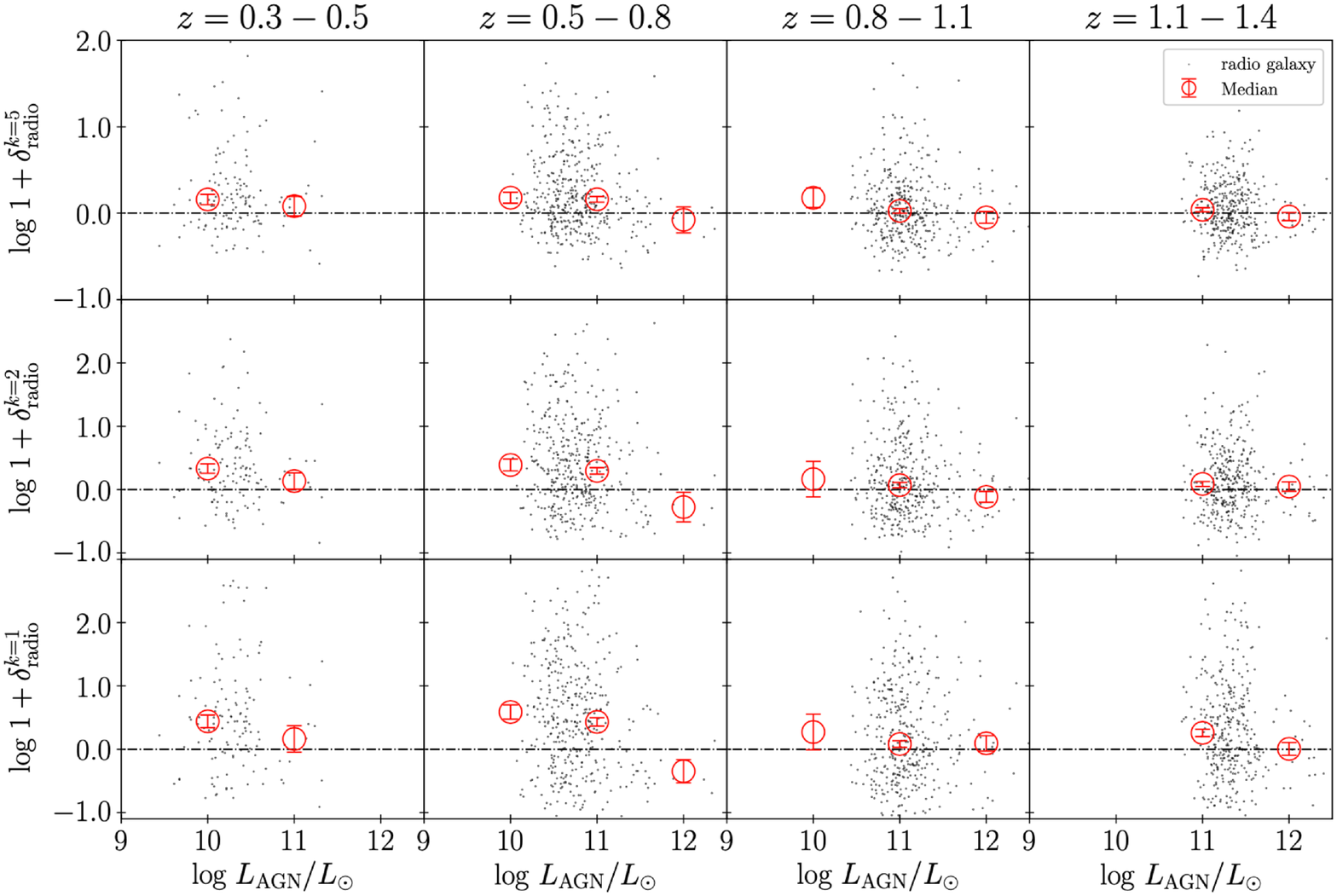}
\end{center}
\caption{
The dependence of the overdensities of the radio galaxies on the $L_{\mathrm{AGN}}$ in a logarithmic scale at $z=0.3-0.5$ (left panels), $z=0.5-0.8$ (middle left panels), $z=0.8-1.1$ (middle right panels), and $z=1.1-1.4$ (right panels). 
Bottom, middle, and top panels show the overdensities with $k=1$, 2 and 5, respectively. The median and the standard error of median of the overdensities in each redshift bin are shown by the red open circle and error bar, respectively. 
}\label{lagn_bin}
\end{figure*}

\begin{figure*}
\begin{center}
\includegraphics[width=0.75\linewidth]{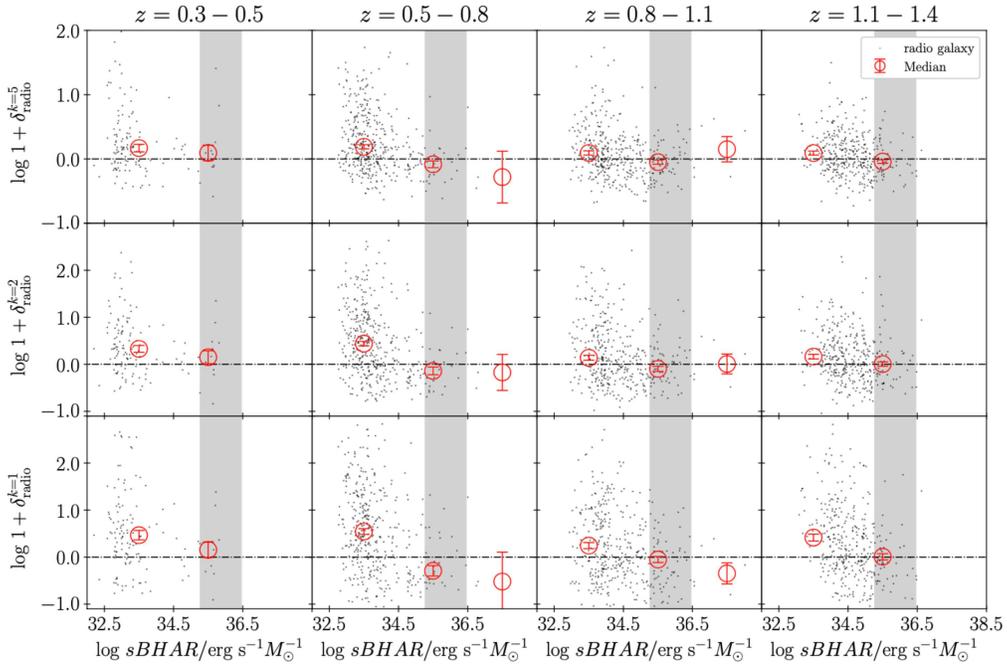}
\end{center}
\caption{Identical to Figure \ref{lagn_bin} but for dependence on the $sBHAR$  in a logarithmic scale. 
The gray shaded regions show the expected $sBHAR$ with $\lambda_{\mathrm{Edd}} = 1$ and $M_*=10^{9}-10^{12}M_\odot$, estimated through equation (\ref{sbhar_eq}). 
}\label{sbhar_bin}
\end{figure*}

\begin{figure*}
\begin{center}
\includegraphics[width=0.75\linewidth]{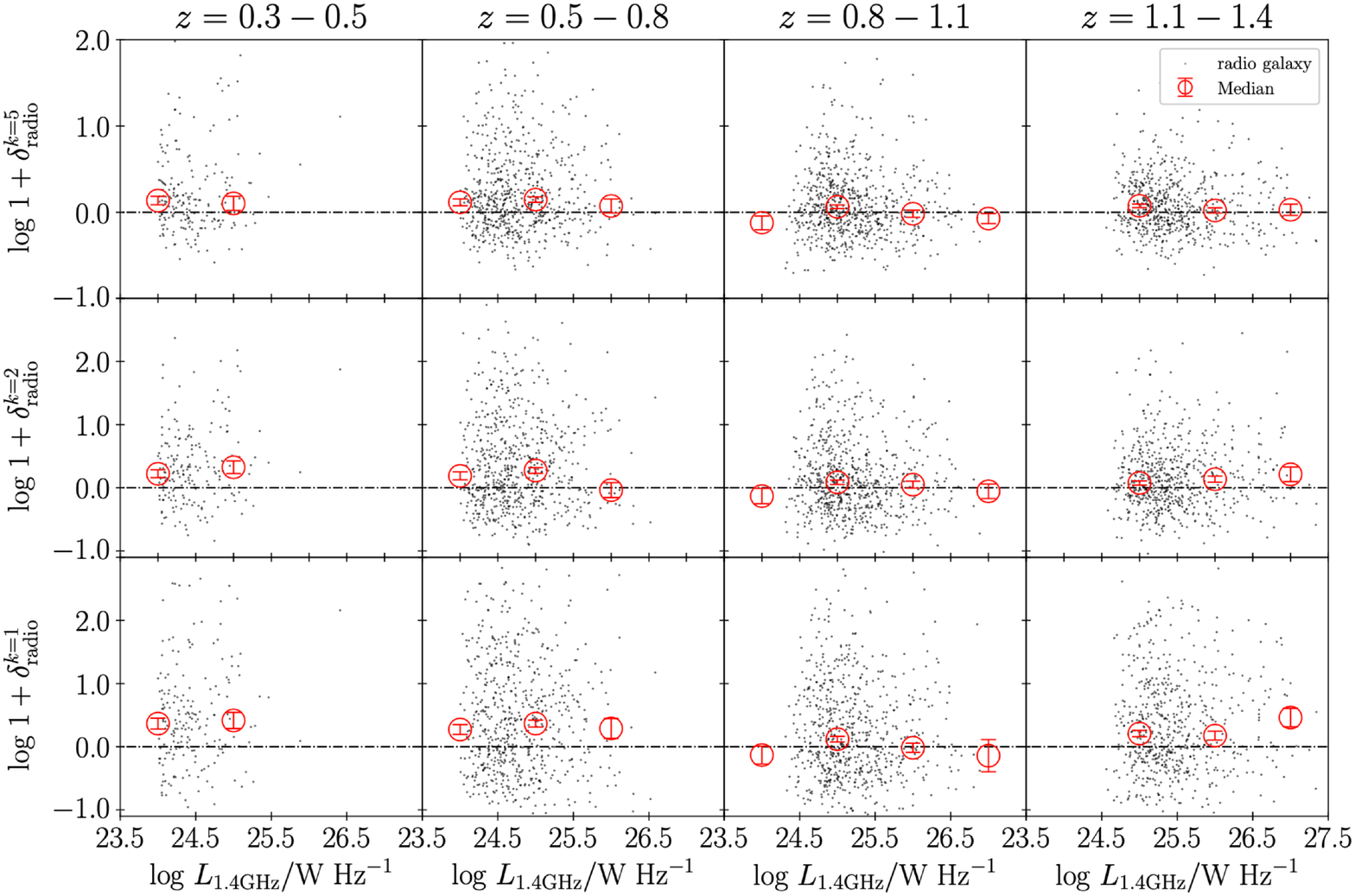}
\end{center}
\caption{Identical to Figure \ref{lagn_bin} but for dependency on the $L_{1.4 \mathrm{GHz}}$  in a logarithmic scale. 
}\label{radiolum_bin}
\end{figure*}

\begin{figure*}
\begin{center}
\includegraphics[width=0.75\linewidth]{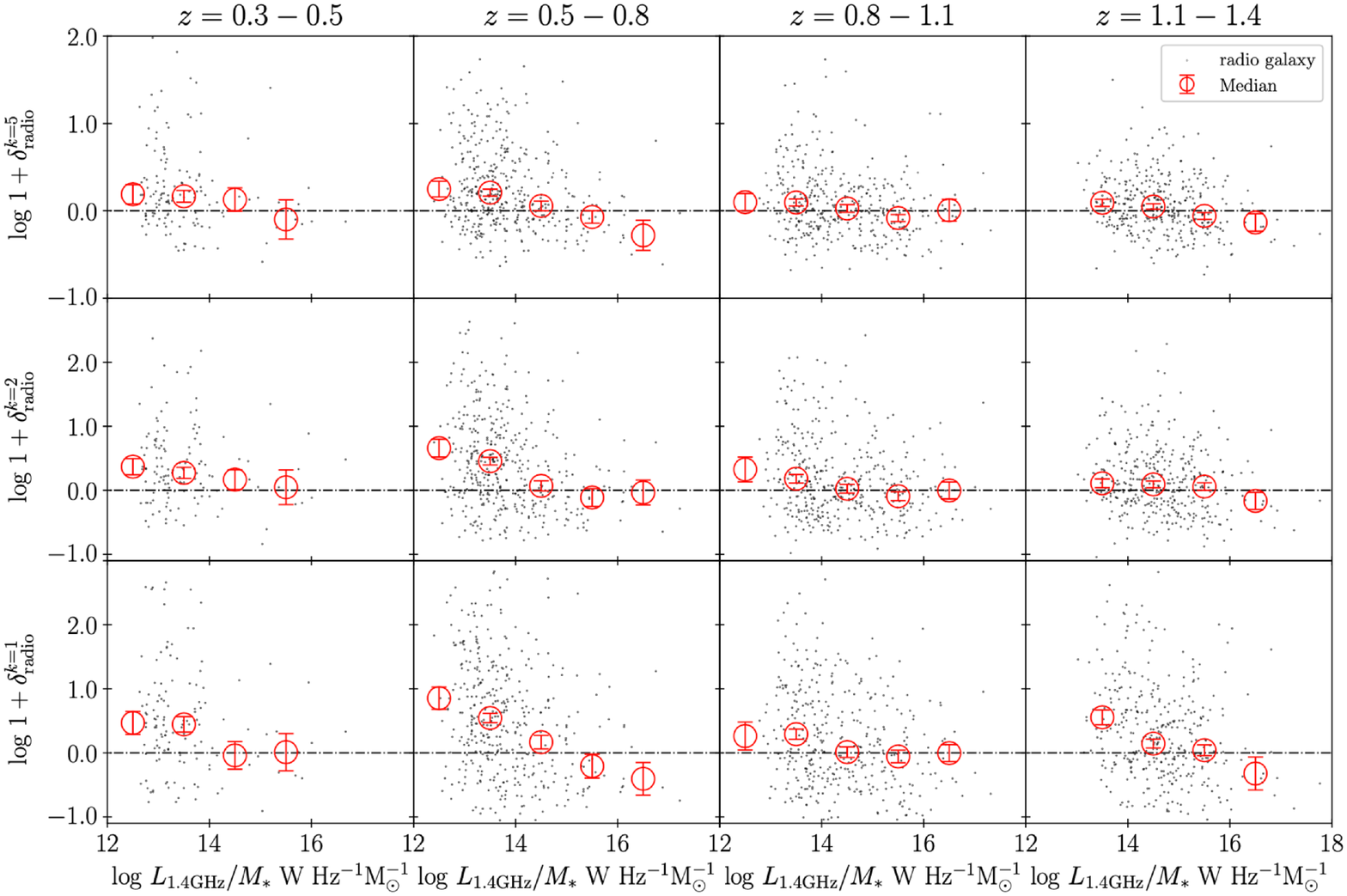}
\end{center}
\caption{Identical to Figure \ref{lagn_bin} but for dependency on the $L_{1.4 \mathrm{GHz}}/M_*$  in a logarithmic scale. 
}\label{r_bin}
\end{figure*}

\begin{deluxetable*}{@{\extracolsep{4pt}}llrlrlrlrlrlrlrlrlr}
\tablecaption{Spearman rank correlation test for overdensity and AGN state at each redshift bin. \label{t_agn_bin}}
\tablecolumns{9}
\tablewidth{0pt}
\tablehead{
\colhead{}   & \multicolumn{2}{c}{$z=0.3-0.5$} & \multicolumn{2}{c}{$z=0.5-0.8$} & \multicolumn{2}{c}{$z=0.8-1.1$}& \multicolumn{2}{c}{$z=1.1-1.4$}   \\
\colhead{overdensity}    & \colhead{$\rho$ \tablenotemark{a}}     & \colhead{$P$\tablenotemark{b}}  & \colhead{$\rho$\tablenotemark{a}}     & \colhead{$P$\tablenotemark{b}} & \colhead{$\rho$ \tablenotemark{a}}     & \colhead{$P$\tablenotemark{b}}  & \colhead{$\rho$\tablenotemark{a}}     & \colhead{$P$\tablenotemark{b}}
}
\startdata  
$L_{\mathrm{AGN}}$ \\
\hline
$1+\delta^{k=1}$          &$0.017$                           & $0.85$                  &$-0.11$                           &$0.030$   &$-0.031$                           & $0.55$                  &$-0.045$                           &$0.40$         \\ 
$1+\delta^{k=2}$          &$-0.035$                           & $0.69$                  &$-0.095$                           &$0.056$   &$-0.058$                           & $0.26$                  &$-9.2\times10^{-3}$                           &$0.86$          \\ 
$1+\delta^{k=5}$         &$0.012$                          & 0.89                  &$-0.12$                           &$0.019$   &$-0.059$                           & $0.25$                  &$-0.040$                           &$0.44$          \\
\hline
$sBHAR$ \\
\hline
$1+\delta^{k=1}$          &$-0.17$                           & $0.056$                  &$-0.32$                           &$1.4\times10^{-10}$   &$-0.23$                           & $7.7\times10^{-6}$                  &$-0.33$                           &$1.1\times10^{-10}$         \\ 
$1+\delta^{k=2}$          &$-0.21$                           & $0.017$                  &$-0.33$                           &$7.8\times10^{-12}$   &$-0.24$                           & $2.3\times10^{-6}$                  &$-0.23$                           &$8.7\times10^{-6}$          \\ 
$1+\delta^{k=5}$         &$-0.19$                          & 0.027                  &$-0.30$                           &$4.9\times10^{-10}$   &$-0.25$                           & $9.7\times10^{-7}$                  &$-0.22$                           &$2.3\times10^{-5}$       \\
\hline
$L_{1.4\mathrm{GHz}}$ \\
\hline
$1+\delta^{k=1}$          &0.087                           & $0.23$                  &$-0.016$                           &$0.70$   &$-0.052$                           & $0.18$                  &0.023                           &$0.58$         \\ 
$1+\delta^{k=2}$          &0.12                           & $0.11$                  &$2.6\times10^{-3}$                           &$0.95$   &$-0.071$                           & 0.061                  &0.028                           &$0.48$          \\ 
$1+\delta^{k=5}$         &$-0.023$                          & $0.75$                  &0.023                           &$0.56$   &$-0.067$                           & $0.080$                  &$-0.075$                           &$0.066$         \\
\hline
$L_{1.4\mathrm{GHz}}/M_*$  \\
\hline
 $1+\delta^{k=1}$          &$-0.13$                           & $0.15$                  &$-0.27$                           &$3.7\times10^{-8}$   &$-0.23$                           & $8.2\times10^{-6}$                  &$-0.26$                           &$3.8\times10^{-7}$         \\ 
$1+\delta^{k=2}$          &$-0.080$                           & $0.36$                  &$-0.27$                           &$5.3\times10^{-8}$   &$-0.21$                           & $2.6\times10^{-5}$                  &$-0.20$                           &$1.3\times10^{-3}$          \\ 
$1+\delta^{k=5}$         &$-0.15$                          & $0.024$                  &$-0.18$                           &$1.9\times10^{-4}$   &$-0.21$                           & $4.2\times10^{-5}$                  &$-0.20$                           &$9.4\times10^{-5}$   
\enddata
\tablenotetext{a}{Correlation coefficient of Spearman rank correlation test. }
\tablenotetext{b}{$P$-value of Spearman rank correlation test. }
\end{deluxetable*}

\subsection{Dependence of the density environments on the SMBH states of radio galaxies}  

\citet{Ching17} reported that HERGs tend to reside in the lower density regions compared to LERGs at $z<0.4$. 
If this result is valid even at $z=0.3-1.4$, this implies that the ambient densities of the radio galaxies are anti-correlated with the radio luminosities and the mass accretions onto their SMBHs, because HERGs have high SMBH accretion rates and radio luminosities compared to LERGs \citep[e.g.,][]{Heckman14, Miraghaei17}.  
In this subsection, we discuss the possible correlation between the ambient environments and such AGN states of radio galaxies at higher-$z$. 

Figure \ref{lagn_bin} shows  the overdensity of the IR radio galaxies as a function of the bolometric AGN luminosity $L_{\mathrm{AGN}}$. 
Objects with higher $L_{\mathrm{AGN}}$ have higher mass accretion on their SMBHs. 
The overdensities with all $k$, on average, tend to be higher than unity at the faint-end regime of the AGN luminosities. 
As the AGN luminosities increase, the overdensities decrease and eventually reach around unity at all the redshift bins. 
However, these relations are not significant according to the Spearman rank correlation test (Table \ref{t_agn_bin}).  

Note that the AGN luminosities strongly depends on the black hole masses \citep[e.g.,][]{Woo02} and thus, the stellar masses. 
In order to capture the mass accretion rate onto SMBH, we should use the AGN luminosities normalized with the stellar masses (that is, $sBHAR$). 
Figure \ref{sbhar_bin} shows the relation between the overdensities and $sBHAR$ of the IR radio galaxies. 
Both are negatively correlated with each other. 
The weak but significant negative correlation between the $sBHAR$ and overdensities is guaranteed by the result of the Spearman rank correlation test (Table \ref{t_agn_bin}).  
At the low $sBHAR$ regime, the overdensities are, on average, significantly higher than unity. 
On the other hand,  at the high $sBHAR$ regime, the overdensities of the IR radio galaxies are comparable to be around unity. 
The fraction of the IR radio galaxies with $\lambda_{\mathrm{Edd}}\ge1$ tend to increase with redshift ($\sim8$ \% at $z<0.8$, and $\sim15$ \% at $z>0.8$). 


We further examine the dependence of the overdensities on the radio luminosities, $L_{1.4\mathrm{GHz}}$, which can trace the radio jet power  \citep[e.g.,][]{Fanidakis11}. 
As shown in Figure \ref{radiolum_bin}, no correlation between the overdensities and radio luminosities are found. 
No significant correlation between them is also found by the Spearman rank correlation test  (Table \ref{t_agn_bin}). 
This is consistent with the result of \citet{Kolwa19} who found no correlation between the surrounding densities and radio luminosities of radio galaxies at $z<0.8$.

As is the case of the AGN luminosity, the radio luminosity depends on the black hole mass and thus stellar mass \citep[e.g.,][]{Fanidakis11}. 
In order to separate HERGs and LERGs in the radio luminosity space more effectively, 
the radio luminosity should be normalized with the stellar mass \citep[e.g.,][]{Miraghaei17}.  
Figure \ref{r_bin} shows the relation between the overdensities and normalized radio luminosities $L_{1.4\mathrm{GHz}}/M_*$ [W Hz$^{-1}$ $M_\odot^{-1}$] of the radio galaxies. 
The radio galaxies with the faint-end normalized radio luminosities, on average, tend to reside in the overdense regions. 
The surrounding densities of the radio galaxies tend to decrease as the normalized radio luminosities increase. 
The local overdensities of the bright-end radio galaxies converge to unity or less than unity. 
The Spearman rank correlation test proposes a significant negative correlation between the normalized radio luminosities and the overdensities of the radio galaxies as summarized in Table \ref{t_agn_bin}. 

To summarize, the radio galaxies with low mass accretion and low radio luminosity tend to reside in galaxy overdensity regions, on average, at fixed stellar mass. 
These findings are qualitatively consistent with the result at $z<0.3$ found by \citet{Ching17} that LERGs reside in the higher density regions compared to HERGs. 
Our results suggest that the correlation of the SMBHs and ambient densities of radio galaxies are guaranteed up to $z=1.4$. 
The anti-correlation between the radio jet luminosities and surrounding densities of the radio galaxies is also consistent with the theoretical model of radio jets proposed by \citet{Kawakatu09}, if the core matter densities of the radio galaxies are correlated with the surrounding galaxy densities. 
This model shows that if the matter densities in the core of the radio galaxies are relatively high, the radio jets launched from the radio galaxies cannot expand beyond host halo due to the energy loss of the radio jets by the interaction with the surrounding matters. This would result in the LERG-like radio galaxies.


\subsection{Triggering of radio galaxy and role of the local environment}  
Triggering of radio galaxy is closely associated with the local density environment \citep[e.g.,][]{Kolwa19}. 
Especially, the distance to the nearest neighbor can be a good measure to diagnose whether radio galaxy triggering is linked with galaxy mergers or secular process \citep[][]{Ching17}. 
In this subsection, we examine if  $d_{k=1}$ of the radio galaxies are typical major merger scale, and discuss whether the radio galaxies are powered by mergers.  

\begin{figure}
\begin{center}
\includegraphics[width=1.0\linewidth]{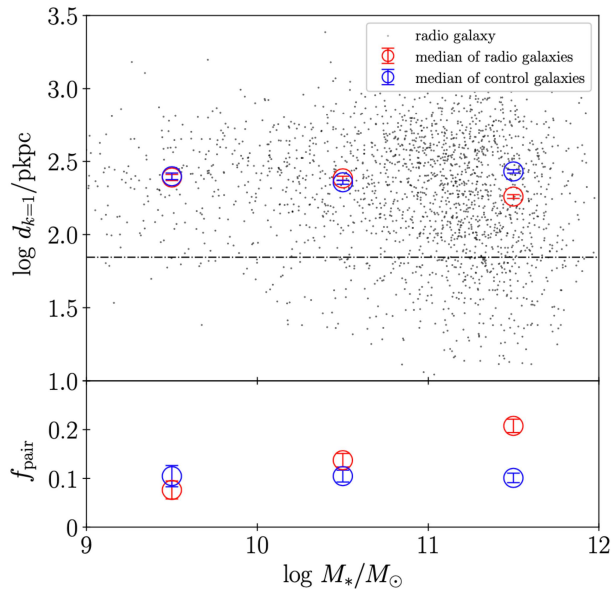}
\end{center}
\caption{The projected distances from the radio galaxies (red points) and control galaxies (blue points) to the nearest neighbors (top panel) and their pair fractions (bottom panel) as a function of stellar mass.   
The median and standard error of median of the projected distances and pair fractions in each stellar mass bin in a logarithmic scale are shown by the  open circle and error bar, respectively. 
}\label{distance}
\end{figure}

In section 4.2, we found that 
the massive radio galaxies with log $M_*/M_\odot > 11$ reside in high density regions compared to control galaxies. 
The density enhancements around radio galaxies are most prominent when $k = 1$. 
These facts suggest that the projected distances from the radio galaxies to the nearest neighbors, $d_{k=1}$, are, on average, small compared to the control galaxies at the massive-end. 
Upper panel in Figure \ref{distance} shows the dependence of  $d_{k=1}$ of the radio and control galaxies on the stellar masses. 
We find that the median $d_{k=1}$ of the radio galaxies is significantly smaller than that of the control galaxies at the stellar mass regime of log $M_*/M_\odot > 11$. 

We define the pair fraction, $f_{\mathrm{pair}} (M_*)= n_{d<70}(M_*)/n_{\mathrm{tot}}(M_*)$, where $n_{d<70}(M_*)$ is the number of the radio/control galaxies with $d_{k=1}<70$ pkpc, which corresponds to the separation of two galaxies at the beginning of a major merger \citep[][]{Larson16}, 
and  $n_{\mathrm{tot}}(M_*)$ is the total number of the radio or control galaxies in a given stellar mass regime of log $M_*\pm0.5$. 
Lower panel in Figure \ref{distance} shows the pair fraction for the stellar masses of the radio and control galaxies. 
The pair fraction is estimated to be $\sim0.1-0.2$ which is comparable to the results of previous studies at this epoch \citep[e.g.][]{Lotz11}. 
We find that the pair fraction of radio galaxies is significantly higher than that of the control galaxies at log $M_*/M_\odot>10$. 
This suggests that the triggering of a significant fraction of massive-end radio galaxies is associated with galaxy mergers. 
The values of  $d_{k=1}$ and $f_\mathrm{pair} (M_*)$ for each stellar mass are summarized in Table \ref{t_merger_bin}.



Less massive radio galaxies appear to have a different triggering mechanism to the higher mass radio galaxies, 
because in log $M_*/M_\odot < 11$ the pair fraction and $d_{k=1}$ (or local densities) of the less-massive radio galaxies are comparable to those of the control galaxies within $1\sigma$ error (Figure \ref{distance}, and section 4.2). 
We also found that in section 5.2, the radio galaxies with high black hole accretion rates tend to reside in the low-density regions.  
These findings imply that high mass accretions onto SMBHs of less-massive radio galaxies occur, regardless of the richness of the associated environments. 
This is consistent with a recent result of \citet{Davis22} who studied radio AGN hosted by dwarf galaxies at $z<0.5$ by combining HSC and Low-Frequency Array (LOFAR) data. 
\citet{Davis22} found that the ignitions of AGN in dwarf galaxies come from self-interactions, not from interactions with its surroundings. 

Our findings imply that massive radio galaxies have experienced galaxy mergers in the past,  have already grown up by $z>1.4$, and then at $z=0.3-1.4$, the mass accretions onto SMBHs have almost ceased, because galaxy merger rate is expected to increase with stellar mass and redshift  \citep[e.g.,][]{Hopkins10}. 
Less-massive radio galaxies, on the other hand, are expected to have avoided merger events and are beginning to undergo active accretion just at $z=0.3-1.4$.  
According to recent simulations \citep[e.g.,][]{Bower17,Habouzit17}, the reservoir gas stored in less-massive galaxy is expected to be easily blown off and carried beyond the host halo by star formation pressures. 
Such negative feedback prevents the gas from falling into the central SMBH, and causes ``galaxy  grows first, BH comes later" phase \citep[e.g., ][]{Ichikawa21}. 
This phase is expected to continue until the stellar mass reaches critical one \citep[$M_*\sim10^{10.5}M_\odot$; ][]{Bower17,Habouzit17}. 
Once the stellar mass approaches to critical one, the negative feedback no longer has enough energy to blow the gas beyond the host halo.  
Then, the active mass accretion to the central SMBH can begin. 
Most of the less-massive radio galaxies are expected to have been spared from the environmental effects that promote the rapid growth phase of SMBHs, such as galaxy mergers. 


\begin{deluxetable*}{@{\extracolsep{4pt}}llrlrlrlrlrlrlrlrlr}
\tablecaption{$d_{k=1}$ and $f_{\mathrm{pair}}$ at each stellar mass bin. \label{t_merger_bin}}
\tablecolumns{9}
\tablewidth{0pt}
\tablehead{
\colhead{}   & \multicolumn{4}{c}{radio galaxy} & \multicolumn{4}{c}{control galaxy}   \\
\cline{2-5}  \cline{6-9} 
\colhead{}    & \colhead{$10^{8-9}M_{\odot}$}     & \colhead{$10^{9-10}M_{\odot}$}     & \colhead{$10^{10-11}M_{\odot}$} & \colhead{$10^{11-12}M_{\odot}$}     & \colhead{$10^{8-9}M_{\odot}$}  & \colhead{$10^{9-10}M_{\odot}$}     & \colhead{$10^{10-11}M_{\odot}$} & \colhead{$10^{11-12}M_{\odot}$} 
}
\startdata 
$d_{k=1}$ [pkpc]         & $212.3\pm24.6$   & $245.4\pm12.1$    & $242.2\pm8.3$  & $181.9\pm5.6$  
                                   &$231.1\pm36.4$    & $250.2\pm13.2$   & $227.7\pm7.3$   & $270.1\pm7.3$         \\ 
$f_{\mathrm{pair}}$     & $0.12\pm0.05$     &$0.08\pm0.02$       & $0.14\pm0.01$   &$0.21\pm0.01$   
                                  &$0.13\pm0.06$       & $0.10\pm0.02$     &$0.10\pm0.01$    &$0.10\pm0.01$          \\ 
\enddata
\end{deluxetable*}

\section{Conclusion} 
We embraced the wide and deep imaging data of HSC-SSP to characterize the redshift evolution of radio galaxy environments between  $z=0.3-1.4$. 
The radio galaxy sample was extracted from the WERGS data \citep[][]{Yamashita18}. 
The redshifts and stellar masses are estimated by Mizuki SED fitting code \citep[][]{Tanaka18}. 
The control sample was constructed by matching to the stellar masses and redshifts of the radio galaxies. 
Multi wavelength data from \citet{Toba19}, for the radio galaxies was used to examine 
the possible correlation between the surrounding density environments and the AGN states of the radio galaxies.  
In order to define the surrounding densities around the radio and control galaxies, the $k$-nearest neighbor method was used. 
We examined the redshift evolution of the local densities of the radio galaxies, and the possible relations between the densities and the properties such as stellar mass and AGN states of the radio galaxies. 
Our findings are as follows: 
\begin{itemize}

\item 
There is no correlation between the overdensities and redshifts of the control galaxies, while the overdensities of the radio galaxies are significantly but weak anti-correlated with redshift. 

\item 
In the low stellar mass regime of log $M_*/M_\odot < 11$, both of the radio and control galaxies, on average, reside in similar environments with the overdensities of $\sim1$. 
On the other hand, at log $M_*/M_\odot > 11$, 
the control galaxies tend to exist in regions with the overdensities higher than unity, while the median density of the radio galaxies is significantly higher than that of the control galaxies. 


\item
The projected distances from the radio galaxies to their nearest neighbors are significantly smaller than those of the control galaxies at the massive-end. 
On the other hand, no difference in the projected distances is found between the radio and control galaxies at the less-massive end. 


\item 
The radio luminosities and $sBHAR$ of the radio galaxies are significantly anti-correlated with their surrounding densities at fixed stellar mass. 

\end{itemize}

Our results support the known scenario where the relative abundance of less-massive HERGs increases with redshift. HERGs tend to reside in low density regions compared to LERGs. 
In addition, our findings suggest that massive radio galaxies have already matured through galaxy mergers in the past and have SMBHs whose mass accretion almost ceased between $z=0.3-1.4$, while less-massive radio galaxies undergo active accretion in this epoch and have avoided any merger events. 


\acknowledgments

\begin{center}
{ ACKNOWLEDGEMENTS } 
\end{center}

We are deeply grateful to the referee for his/her helpful comments that improved the manuscript. 
This work is supported by  Japan Society for the Promotion of Science (JSPS) KAKENHI (22K14075 and 21H04490; HU, 21K13968 and 20H01939; TY, 20H01949: TN). 

This work is based on data collected at the Subaru Telescope and retrieved from the HSC data archive system, which is operated by Subaru Telescope and Astronomy Data Center at National Astronomical Observatory of Japan. 

The Hyper Suprime-Cam (HSC) collaboration includes the astronomical communities of Japan and Taiwan, and Princeton University. The HSC instrumentation and software were developed by the National Astronomical Observatory of Japan (NAOJ), the Kavli Institute for the Physics and Mathematics of the Universe (Kavli IPMU), the University of Tokyo, the High Energy Accelerator Research Organization (KEK), the Academia Sinica Institute for Astronomy and Astrophysics in Taiwan (ASIAA), and Princeton University. Funding was contributed by the FIRST program from Japanese Cabinet Office, the Ministry of Education, Culture, Sports, Science and Technology (MEXT), the Japan Society for the Promotion of Science (JSPS), Japan Science and Technology Agency (JST), the Toray Science Foundation, NAOJ, Kavli IPMU, KEK, ASIAA, and Princeton University. 

This paper makes use of software developed for the Large Synoptic Survey Telescope. We thank the LSST Project for making their code available as free software at  http://dm.lsst.org

The Pan-STARRS1 Surveys (PS1) have been made possible through contributions of the Institute for Astronomy, the University of Hawaii, the Pan-STARRS Project Office, the Max-Planck Society and its participating institutes, the Max Planck Institute for Astronomy, Heidelberg and the Max Planck Institute for Extraterrestrial Physics, Garching, The Johns Hopkins University, Durham University, the University of Edinburgh, Queen's University Belfast, the Harvard-Smithsonian Center for Astrophysics, the Las Cumbres Observatory Global Telescope Network Incorporated, the National Central University of Taiwan, the Space Telescope Science Institute, the National Aeronautics and Space Administration under Grant No. NNX08AR22G issued through the Planetary Science Division of the NASA Science Mission Directorate, the National Science Foundation under Grant No. AST-1238877, the University of Maryland, and Eotvos Lorand University (ELTE) and the Los Alamos National Laboratory.




\end{document}